\documentclass{elsart}

\usepackage{psfig}
\newcommand{\beq}{\begin{equation}}
\newcommand{\eeq}[1]{\label{#1} \end{equation}}
\newcommand{\insertplotshort}[1]{\centerline{\psfig{figure={#1},height=11.0cm}}}
\newcommand{\insertplotl}[1]{\centerline{\psfig{figure={#1},height=19.0cm}}}

\begin{document}

\begin{frontmatter}



\title{Can supercooling explain the HBT puzzle?}


\author[label1,label2]{L.P. Csernai}
\ead{csernai@fi.uib.no}
\author[label3,label4]{M.I. Gorenstein}
\ead{goren@th.physik.umi-frankfurt.de}
\author[label3]{L.L. Jenkovszky}
\ead{jenk@gluk.org}
\author[label5]{I. Lovas}
\ead{lovah@dtp.atomki.hu,lovah@aph.atomki.hu}
\author[label3,label6]{V.K. Magas}
\ead{vladimir@cfif.ist.utl.pt}

\address[label1]{
Section for Theoretical and Computational Physics, Department of Physics\\
University of Bergen, Allegaten 55, N-5007, Norway
}
\address[label2]{
KFKI Research Institute for Particle and Nuclear Physics\\
P.O.Box 49, 1525 Budapest, Hungary
}
\address[label3]{Bogolyubov Inst. for Theor. Phys., Ac. of Sciences
of Ukraine\\
 Metrologichna str. 14-b, 03143 Kiev, Ukraine
}
\address[label4]{University of Frankfurt\\ 
Robert-Mayer St. 8-10, D-60054 Frankfurt am Main, Germany
}
\address[label5]{University of Debrecen\\
Egytem-ter 1-3, H-4010 Debrecen, Hungary
}
\address[label6]{
Center for Physics of Fundamental Interactions (CFIF)\\
Instituto Superior Tecnico, Av. Rovisco Pais, 1049-001 Lisbon, Portugal
}

\begin{abstract}
Possible hadronization of supercooled QGP, created in heavy ion collisions at RHIC and SPS,
is discussed within a Bjorken hydrodynamic model. Such a hadronization is expected to be a 
very  fast shock-like process, what, if hadronization coincides or shortly followed by freeze out,
could explain a part of the HBT puzzle, i.e. the flash-like particle
emission ($R_{out}/R_{side}\approx 1$).
HBT data also show that the expansion
time before freeze out is very short ($\sim 6-10\ fm/c$). In this work we discuss
question of supercooled QGP and the timescale of the reaction.
\end{abstract}


\end{frontmatter}

\section {Introduction} \label{s1}

Two-particle interferometry has become a powerful tool for studying the
size and duration of particle production from elementary collisions
($e^+e^-$, $pp$ and $p\bar{p}$) to heavy ions like $Au+Au$ at RHIC or
$Pb+Pb$ at SPS ~\cite{HBT,BertschPratt}. For the case of nuclear
collisions, the interest mainly focuses on the possible transient
formation of a deconfined state of matter. This could affect the size of
the region from where the hadrons (mostly pions) are emitted as well as
the time for particle production.

Comparing recent data~\cite{data} from RHIC with SPS data one finds a
``puzzle''~\cite{Gyulassy:2001zv}: all the HBT radii are pretty similar
although the center of mass energy is changed by an order of magnitude.
Discussions at "Quark Matter 2002" \cite{QMtalks} 
lead to the conclusion that the
duration of particle emission, as well as
the lifetime of the system before freeze out,
appear to be shorter than the predictions of most of the model at
the physics market.

It was demonstrated that a strong first-order QCD phase transition within
continuous hydrodynamical expansion
would lead to long lifetimes of
the particle source~\cite{BertschPratt,RiGy,TLS}\footnote{
If we use some microscopic model for
hadronization, for example nucleation of relativistic first-order phase
transition \cite{CsKa92}, the lifetime is even longer - it was estimated
to be about $50-100\ fm/c$ in Ref. \cite{CsKa92}.}, which would
manifest itself
as a large $R_{out}/R_{side}$ ratio.
Now this type of hadronization is excluded by experimental data.

An alternative possibility, discussed in  Refs.
\cite{CsCsorgo,CsMish,goren,Antti,eos3}, is the hadronization from the
supercooled QGP. This is expected to be a very fast
shock-like process. If the hadronization from supercooled QGP
coincides with freeze out, like it was assumed in Ref. \cite{Antti},
then this could explain a part of the HBT puzzle,
i.e. the flash-like particle
emission ($R_{out}/R_{side}\approx 1$).
In this work we are asking the following question -- can the
hadronization from supercooled QGP explain also the another part of the 
HBT puzzle, i.e. a very short
($\sim 6$\cite{Csorgo} $-\ 10$\cite{QMtalks,HBTtalk} $fm/c$) expansion
time before freeze out?

\section{Shock hadronization of a sQGP} \label{shock}

Relativistic shock phenomena were widely discussed with respect to
their
connection to high-energy heavy ion collisions (see, for example,
\cite{r1}). In thermal equilibrium by admitting the existence of the
sQGP and the superheated
hadronic matter (HM) we have essentially richer picture of
discontinuity-like transitions than in standard compression and
rarefaction shocks.
The system evolution in relativistic hydrodynamics is governed by the
energy-momentum tensor $ T^{\mu\nu} = (\epsilon + p)u^{\mu}u^{\nu} -
pg^{\mu\nu}$ and conserved charge currents (in our applications to heavy
ion collisions we consider only the baryonic current $ nu^{\mu}$). They
consist of local thermodynamical fluid quantities (the energy density
$\epsilon$, pressure $p$, baryonic density $n$) and the collective
four-velocity $u^{\mu} = \sqrt{1-{\bf v}^{2}}(1,{\bf v})$. Continuous
flows are the solutions of the hydrodynamical equations:
 \beq
\partial_{\mu} T^{\mu\nu} = 0~,~~
\partial_{\mu} nu^{\mu} = 0 ~,
\eeq{e02}
with specified initial and boundary conditions. These
equations are nothing more than the differential form of the
energy-momentum and baryonic number conservation laws. Along with
these continuous flows, the conservation laws can also be realized
in the form of discontinuous hydrodynamical flows which are called
shock waves and satisfy the following equations: \beq
 T_o^{\mu\nu}d\sigma_{\nu} =
 T^{\mu\nu}d\sigma_{\nu}  \;\; ,~~
 n_{o}u_{o}^{\mu}d\sigma_{\mu}=
 n u^{\mu}d\sigma_{\mu} \;\; ,
\eeq{e12} where $d\sigma ^{\mu}$ is the unit 4-vector normal to the
discontinuity hypersurface. In eq. (\ref{e12}) the zero index corresponds
to the initial state ahead of the shock front and quantities without an
index are the final state values behind it. A general derivation of the
shock equations (valid for both space-like and time-like normal vectors
$d\sigma ^{\mu}$) was given
in Ref.~\cite{r6}.

\begin{figure}[htb]
        \insertplotshort{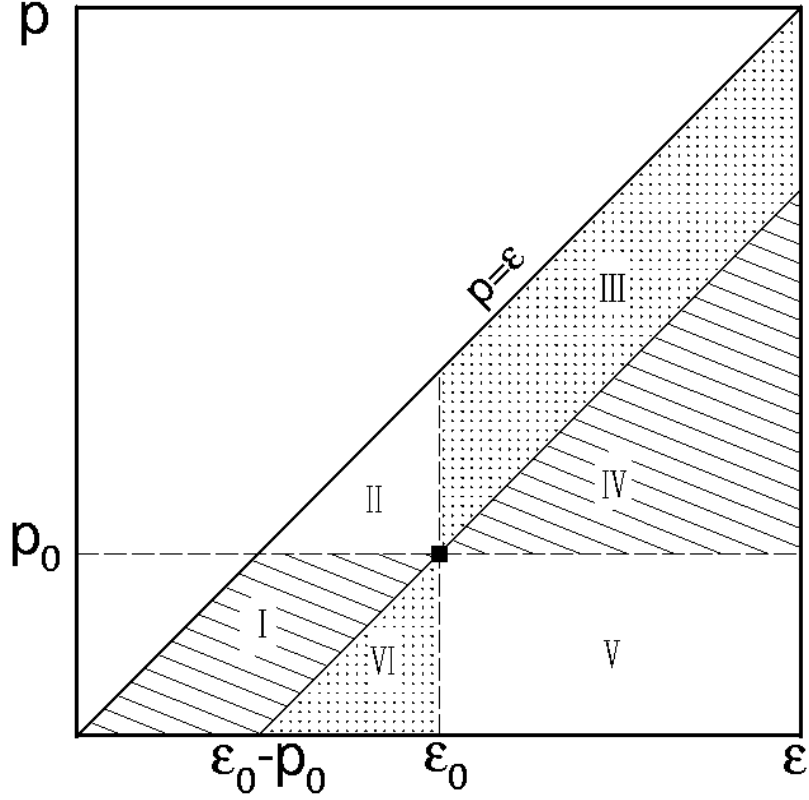}
\caption{ Possible final states in the (energy density--pressure)-plane
for shock transitions from the initial state $(\epsilon_{o},p_{o})$. I and
IV are the physical regions for s.l. shocks, III and VI for t.l. shocks.
II and V are unphysical regions for both types of shocks. Note, that only
states with $p \leq \epsilon$ are possible for any physical Equation of
State in the relativistic theory. }
\label{f1}
\end{figure}

The important constraint on the transitions (\ref{e12}) (thermodynamical
stability condition) is the requirement of non- decreasing entropy ($s$ is
the entropy density):
\beq su^{\mu}d\sigma _{\mu} \ge
s_{o}u_{o}^{\mu}d\sigma _{\mu} \; . \eeq{e2}

To simplify our consideration and make our arguments more transparent we
consider only one-dimensional hydrodynamical motion. To study the shock
transitions at the surface with space-like (s.l.) normal vector (we call
them s.l. shocks) one can always choose the Lorentz frame where the shock
front is at rest. Then $d\sigma ^{\mu} =(0,1)$ at the surface of shock
discontinuity, and eq. (\ref{e12}) in this (standard) case becomes: \beq
T_{o}^{01} = T^{01}, \; \; T_{o}^{11} = T^{11} \;\; ,~~ n_{o}u_{o}^{1} =
nu^{1} \; \; . \eeq{e32} Solving eq. (\ref{e32}) one obtains
\beq v_{o}^{2} = \frac{(p - p_{o})(\epsilon + p_{o})} {(\epsilon -
\epsilon_{o})(\epsilon_{o} + p)} \; , \; \; v^{2} = \frac{(p -
p_{o})(\epsilon_{o} + p)} {(\epsilon - \epsilon_{o})(\epsilon + p_{o})}
\;\; , \eeq{e6} and the well known Taub adiabat (TA) \cite{r8}
\beq n^{2}X^{2} - n_{o}^{2}X_{o}^{2} - (p - p_{o}) (X + X_{o}) = 0 \;\; ,
\eeq{e7} where $X \equiv (\epsilon + p)/n^{2}$.\footnote{It has 
been shown in a series of works \cite{FO}, that freeze out through the 
space-like hypersurface leads to nonequilibrium post FO distribution.}

For discontinuities on a hypersurface with a time-like (t.l.) normal
vector $d\sigma^{\mu}$ (we call them t.l. shocks) one can always choose
another convenient Lorentz frame (``simultaneous system'') where
$d\sigma^{\mu}=(1,0)$. Equation (\ref{e12}) is then
\begin{equation}
T_{o}^{00} = T^{00}~, \; \; T_{o}^{10} = T^{10} \;\; ,~~~
n_{o}u_{o}^{0} = nu^{0} \;\; .\label{e9}
\end{equation}
Solving eq. (\ref{e9}) we find
\begin{equation}
\tilde{v}_{o}^{2} = \frac{(\epsilon - \epsilon_{o})(\epsilon_{o} +
p)} {(p - p_{o})(\epsilon + p_{o})}, \; \; \tilde{v}^{2} = \frac
{(\epsilon - \epsilon_{o})(\epsilon + p_{o})} {(p -
p_{o})(\epsilon_{o} + p)} \;\;,\label{e10}
\end{equation}
where we use the $``\sim"$ sign to distinguish the t.l. shock case (\ref{e10})
from the standard s.l. shocks of (\ref{e6}). Another relation contains
only the thermodynamical variables. It appears to be identical to the TA
of eq. (\ref{e7}).
Eqs. (\ref{e10}) and (\ref{e6}) are connected to each other by simple
relations \cite{goren}:
\begin{equation}
\tilde{v}_{o}^{2} = \frac{1}{v_{o}^{2}}~, \; \; \tilde{v}^{2} = \frac
{1}{v^{2}} \;\; .\label{e_11}
\end{equation}
These relations show that only one kind of transition can be realized for
a given initial state and final state. The physical regions $[0,1)$ for
$v_{o}^{2},v^{2}$ (\ref{e6}) and for $\tilde{v}_{o}^{2},\tilde{v}^{2}$
(\ref{e10}) can be easily found in $(\epsilon $--$p)$-plane \cite{goren}.
For a given initial state $(\epsilon _{o},p_{o})$ they are shown in Fig.
\ref{f1}.
For supercooled initial QGP states the TA no longer passes through the
point $(\epsilon _{o},p_{o})$ and new possibilities of t.l. shock
hadronization transitions to regions III and VI in Fig. \ref{f1} appear.

\section {Hadronization of the sQGP
within Bjorken hydrodynamics}
\label{bjor}

\begin{figure}[htb]
\vspace*{-1.5cm}
        \insertplotl{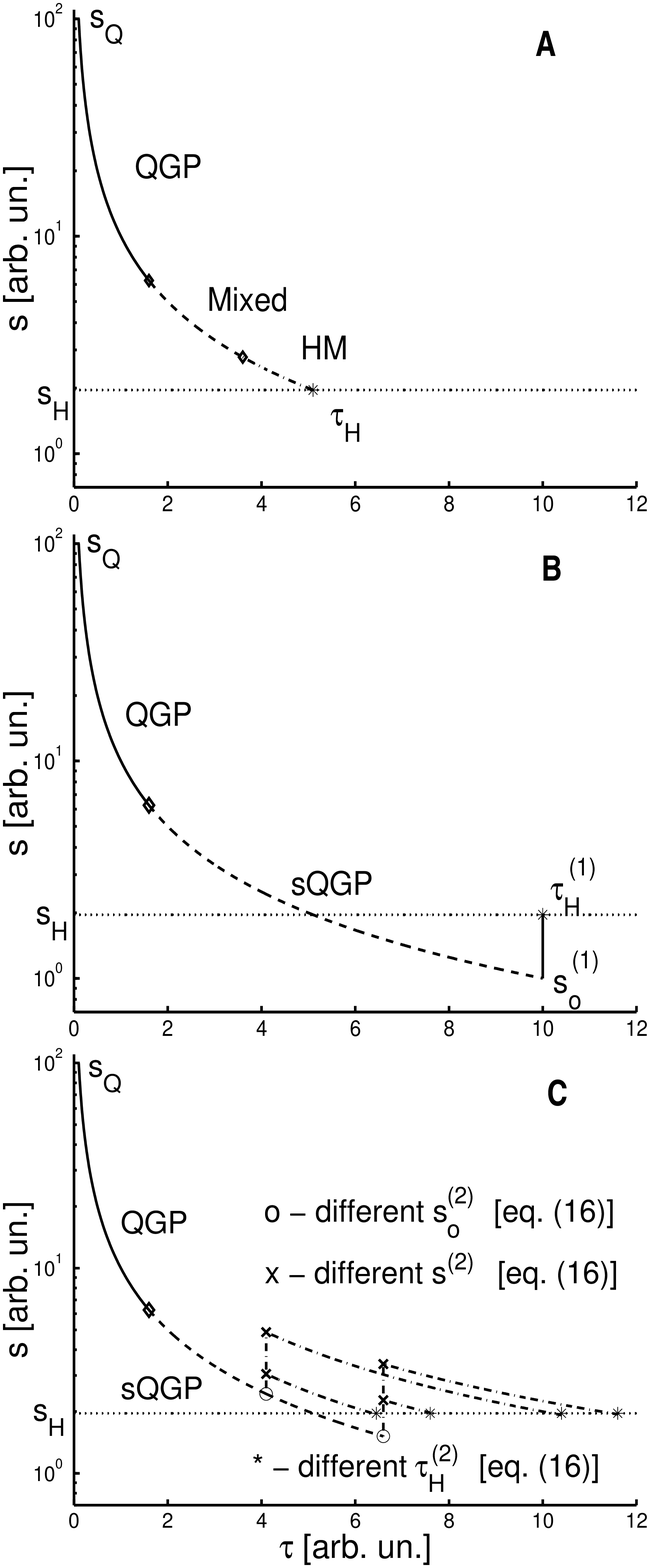}
\vspace*{-0.5cm}
\caption{Different ways for a system to go from Q state ($s_Q$) 
to H state ($s_H$) are presented on $\{s,\tau\}$ plane. 
Subplot A shows continuous expansion, which takes time $\tau_H$, eq. (\ref{t1}). Subplot B presents 
flash-like particle emission, i.e. simultaneous hadronization and freeze out; which takes time $\tau_H^{(1)}$, 
eq. (\ref{t2}). Subplot C shows several possibilities according to scenario 2 
with shock-like hadronization into superheated HM. Time $\tau_H^{(2)}$ (\ref{t3}) 
can  be smaller or  larger than $\tau_H^{(1)}$, 
depending on details of the EoS, but always larger than $\tau_H$.}
\label{f2}
\end{figure}

For a study of the expanding QGP we have chosen a framework of
the one dimensional Bjorken model \cite{r12} 
(actually our principal results will not 
change if we use 3D Bjorken model).
Within
the Bjorken model  all the thermodynamical
quantities are constant along constant proper time curves,
$\tau=\sqrt{t^2-z^2}=const$. The important
result of Bjorken hydrodynamics (which assumes a perfect fluid) 
is that the evolution of the entropy
density, is independent of the Equation of State (EoS),
namely \beq s(\tau)={s(\tau_{init})\tau_{init} \over \tau}\,. \eeq{ent}
In Bjorken model the natural choice of the freeze out hypersurface is
$\tau=const$ hypersurface, where
normal vector is parallel to the
Bjorken flow velocity, $v=z/t$.
Thus, $d\sigma^{\nu} = (1,0)$ in the rest
frames of each fluid element.
This leads to the simple solution of the t.l. shock
equations (\ref{e9}):
\begin{equation}
  \tilde{v}^{2}=\tilde
{v}_{o}^{2} = 0\;, \; \; \epsilon = \epsilon_{o}, \; n = n_{o},
\;\; p \neq p_{o}\;. \label{e14}
\end{equation}
The entropy condition (\ref{e2}) is reduced to \beq s\ge s_o ~.\eeq{e13}

Now let us try to answer the main question of this work -- can QGP
expansion with t.l. shock hadronization of supercooled state be faster
than the hadronization through the mixed phase?
The initial state is given at the proper time $\tau_{init}\equiv \tau_{Q}$,
when the local
thermal equilibrium is achieved in the QGP state $Q\equiv( \epsilon_Q,
p_Q,s_Q)$.
The final
equilibrium hadron state is also fixed, by experiment or otherwise, 
as $H\equiv(\epsilon_H, p_H, s_H)$.
For the continuous expansion given by eq. (\ref{ent}) the proper time
for the $Q\rightarrow H$ transition is
 (see Fig. \ref{f2} - subplot A): 
\beq
\tau_H~=~\frac{s_{Q}\tau_{Q}} {s_H} ~. \eeq{t1}

If our system enters the sQGP phase and the particle emission
is flash-like, i.e. the system hadronizes and freezes out at the same
time, then eq. (\ref{ent}) is also valid all the time with final t.l.
shock transition to the same $H$ state. We call this as a scenario
number one (see Fig. \ref{f2} - subplot A).
Our system should go into supercooled phase to the point where
$\epsilon_{o}^{(1)}=\epsilon_H, n_{o}^{(1)}=n_{H}$, as it is required by
eq. (\ref{e14}).
At this point our sQGP has entropy density
$s_{o}^{(1)}$. It's
value depend on the EoS, but t.l. shock transition is only possible if
$s_{o}^{(1)}\le s_H$ according to eq. (\ref{e13}). Thus, for the proper
time of
$Q\rightarrow H$ transition according to first scenario we have: 
\beq
\tau_H^{(1)}~=~\frac{s_{Q}~\tau_{Q}}{s_{o}^{(1)}}~\ge~ \tau_H \,. \eeq{t2}

We can also study a scenario number two when our system
supercools to the state
$(\epsilon_{o}^{(2)}, p_{o}^{(2)}, s_{o}^{(2)})$, 
then hadronizes to a
superheated HM state
$(\epsilon^{(2)},p^{(2)},s^{(2)})$, 
and then this HM state expands to the
same freeze out 
state $H\equiv(\epsilon_{H},p_{H},s_{H})$.
(see Fig. \ref{f2} - subplot C). 
At the point of the shock transition
one has:
\beq \tau_{o}^{(2)}~=~\frac{s_Q~\tau_{Q}}{s_{o}^{(2)}} \, , \eeq{t3a} 
Then we have
a t.l. shock transition satisfying eq. (\ref{e14}), and following the HM
branch of the hydrodynamical expansion we find: 
\beq
\tau_H^{(2)}~=~\frac{s^{(2)}~\tau_{o}^{(2)}}{s_H}~=~\frac{s_{Q}\tau_{Q}}{s_H}~
\frac{s^{(2)}}{s_{o}^{(2)}}~\ge~ \tau_H \,, \eeq{t3} 
since $s^{(2)}\ge s_{o}^{(2)}$
due to non-decreasing entropy condition (\ref{e13}).
In this second scenario the value of the of entropy
density $s_o^{(2)}$
of sQGP can be both smaller and larger than HM final value
$s_H$. Depending on details of the EoS the proper time $\tau_H^{(2)}$
(\ref{t3}) of the $Q\rightarrow H$ transition can also be smaller as well
as larger than $\tau_H^{(1)}$ (\ref{t2}), but always larger than $\tau_H$.

\section{Conclusions}
\label{con}

The conclusion of our analysis seems to be a rather general one:
the system's
evolution through
a supercooled phase and time-like shock hadronization can not be shorter
than a continuous expansion within the perfect fluid hydrodynamics
independently of the details of EoS and the parameter values of the
initial, $Q$, and final, $H$, states.
Although in such a way we
may achieve flash-like particle emission, supported by the HBT data, the
expansion time becomes longer, making it harder to reproduce the
experimental HBT radii. 

So, how can we achieve shorter freeze out time than the minimal one
coming from (very fast) 
Bjorken expansion 
via thermal and phase equilibrium?
Any delay in the phase equilibration (see assignment 9 in Ref. \cite{cs94})
or/and any dissipative process in our system 
lead to the entropy production, 
what increases the time needed to reduce entropy density to $s_o\le s_H$ 
(for the flash-like particle emission).

The system may, nevertheless, freeze out and hadronize into a non-equilibrated 
hadron gas well before $\tau_H$. This
is possible e.g. through a dominantly s.l. hypersurface
with non-decreasing entropy condition, 
eq. (\ref{e13}) \cite{FO}, at earlier times 
from a slightly supercooled QGP. 
On the other hand a  dominantly s.l. hypersurface
gives a finite duration of the particle 
emission, making it harder to reproduce 
experimental  $R_{out}/R_{side}$ ratio.

The construction of a full reaction model, which  simultaneously 
describes data on two particle interferometry, hadron spectra and
hadron abundances is a formidable task which is still ahead of us.

\section*{Acknowledgements}
We have benefitted from discussions with I. Mishustin,
L. Satarov and Yu.M. Sinyukov.
The financial support from the Humboldt Foundation and INTAS grant
00-00366 are acknowledged.

\end{document}